**Emergence of Symmetry in a System of Distinguishable but Identical Particles**

Allan Tameshtit

*Ronin Institute, 2108 N St Ste N*
*Sacramento, CA 95816, USA*
*allan.tameshtit@alumni.utoronto.ca*

Instead of the conventional construction of symmetric and antisymmetric states by abruptly projecting with the symmetrizer or antisymmetrizer, this paper investigates rapid but continuous symmetrization via environment-induced decoherence. Density operators are transformed to a symmetric observable via a semigroup map obtained from an interacting Hamiltonian. Likewise, a completely positive dynamical map transforms a completely asymmetric state to either a symmetric or antisymmetric state. The theory is applied to analyse a collision between two identical particles having arbitrary spin and negligible spin interactions. We discuss extensions to larger domains via dynamical maps that are not completely positive.

I. **Introduction**

Two fundamental postulates of quantum mechanics involve abrupt changes arising from the action of projection operators: wave function collapse after a measurement and the construction of physical kets with appropriate symmetry [1]. Wave function collapse refers to the transition that occurs after an eigenvalue of an observable is measured, resulting in the normalized projection onto the eigensubspace associated with that eigenvalue immediately after the measurement. Our understanding of this process is informed by environment-induced decoherence, which suggests that the collapse, although typically occurring on short times scales, may not be instantaneous. Similarly, the construction of a physical ket to describe identical particles involves arbitrarily assigning eigenkets to the individual particles (as though they were not identical) and then obtaining a normalized projection by applying the symmetrizer for bosons or antisymmetrizer for fermions, followed by normalization. This paper explores whether a process similar to environment-induced decoherence can explain this latter construction as a rapid but not discontinuous change.

Notions of distinguishability of identical particles and the relation to the (anti)symmetry of their wave functions are central to this paper and are discussed in [2], [3] and [4]. Identical particles may be distinguishable one moment and indistinguishable the next. In such scenarios, we would expect a transition from a description based on Maxwell-Boltzmann statistics to Bose-Einstein statistics (in the case of bosons) or Fermi-Dirac statistics (in the case of fermions). This perspective has been adopted previously. In Ref. [5], a pair of initially distant particles is deemed to be distinguishable and consequently capable of being described by a product state. "As the particles approach each other, a quantum jump takes place upon particle collision, which erases their distinguishability and projects the two-particle state onto an appropriately (anti-)symmetrized state."

Another reference [6], which is listed in the previous one, proposes that an orbital state determines a map from premolecule spin states to antisymmetric molecule states in a model that may elucidate human cognition.

It is instructive to briefly mention various examples, which will also motivate our work. Unimolecular dissociation of the type $AB_2 \rightarrow A + B_2$ can be studied using fluorescence spectroscopy and molecular beam techniques [7]. For instance, $HeI_2$ and $HeBr_2$ are known complexes that can dissociate to $He + I_2$ and $He + Br_2$. The helium atoms in the two complexes are distinguishable due to their different iodine or bromine neighbours. However, after dissociation, which could be induced suddenly with a laser pulse, the two lone helium atoms can become indistinguishable.

A second example involves two identical particles in a box that are distinguishable due to being separated by an impenetrable wall. If the wall is shattered, then soon enough it is not possible to label the particle that was on one side and the particle that was on the other side of the wall, rendering both particles indistinguishable. (A reverse gedanken experiment starts with two identical and indistinguishable particles in one box. After a separation wall is introduced into the box, there is a finite chance that the final state will consist of two particles that are separated by the wall and therefore distinguishable.) Changes that arise after removing a dividing wall in a box of identical particles have been investigated classically as far back as Gibbs [8] and more recently in the quantum mechanical setting in [9].

Another example involves two particles that can be injected into a container via two injection nozzles separated by a distance $2l$. In the absence of preexisting long-range forces, if the two particles are injected simultaneously with respect to the reference frame of the container, then for a period of at least about $l/c$, the particles are oblivious to each other and therefore during this time the two particles are distinguishable as having been ejected from one or the other nozzle. After this time, this distinction may not be possible.

We may also consider two parallel beams of spin 1/2 particles, one beam consisting of particles in one spin state and the other beam consisting of particles in another orthogonal spin state [1], p.1407 and p.1450. If there are no spin-dependent interactions, the particles maintain their different spins, which may be used as a distinguishing property. However, if at some time the particles of one beam traverse a "spin flipper" that converts their spin to that of the other, the particles become indistinguishable and symmetrization postulates must then be brought to bear.

Finally, we often treat a system of particles as though it were a single composite particle. For example, atoms and nuclei in some contexts are considered particles despite the fact that they are composed of various constituents. Two such composite particles may be identical in every way but for the fact that one may be in an excited state relative to the other (e.g., nuclear isomers). Because of their different internal energies, the two particles have different masses and therefore are distinguishable. The excited particle may interact with its surroundings and emit a photon, decreasing its mass to the value of the other. The two may then become indistinguishable. This transition is caused by an interaction with the environment.

Indistinguishability gives rise to uniquely quantum effects related to symmetries of the wave functions describing the foregoing systems. In traditional approaches, identical particles are described by a symmetric or antisymmetric wave function from the outset; when exchange terms [1] are small, so too are symmetry effects. This principle has been called cluster separability [10]. In other theories, some particles like a deuteron and a nucleon are deemed different when certain isospin commutators approach zero [11]. In this paper, we take a different tack, starting with identical, but distinguishable particles that transition to symmetric or antisymmetric states.

We endeavor to shed light on the emergence of symmetry properties of a density operator by presenting models that give rise to environment-induced symmetrization. This idea exploits the fact that both the symmetrizer (or antisymmetrizer) and the environment generally convert factored states to entangled ones. In this vein, we can mention the effect of permutational symmetry on entanglement that was investigated in Refs. [12] and [13]. Returning to the present study, one aim is to explore whether a density operator describing distinguishable particles can be transformed via a semigroup map to a symmetric operator. Another goal is to determine whether it is possible to use a completely positive dynamical map to transform a state lacking symmetry to either a symmetric or antisymmetric state. We will see in Section IV that a map having such transformational properties will be restricted to a smaller domain than for Maxwell-Boltzmann particles. This is not surprising since even in the usual symmetrization prescription involving the antisymmetrizer $A$ and a normalization constant $N$, the positive map of the density operator $\rho \mapsto NA^\dagger \rho A$ is restricted to a smaller set of otherwise physical density operators due to the Pauli exclusion principle. Along these lines, a completely positive map for a subsystem of indistinguishable fermionic particles was ascertained for a subset of states in Ref. [14].

But first, in the next section, we provide nomenclature and background, much well-known from standard monographs, that we will need for the ensuing work. In Section III (for operators) and Section IV (for states), we investigate interactions that can lead to symmetrization. State transformation will be in terms of an operator-sum representation [15], [16]. In Section V, we examine an elastic collision between two identical particles, which serves as an example of where we can apply the theory developed herein. A discussion of our work in Section VI follows, which contemplates the possibility of expanding the domain of our dynamical maps by allowing the operators in an operator-sum representation to depend on the preimage. This is in line with previous work [17] demonstrating the utility of expressing maps that are not completely positive as an operator-sum representation at the price of having to make the operators therein depend on the preimage. An Appendix is included.

## II. Symmetrization Framework

The word "symmetric" in the context of identical particles has two different meanings depending on whether the term is applied to states or operators. Density operators are special because they can be categorized as both states and operators, and therefore to write that the density operator $\rho$ is symmetric is ambiguous. From a one-particle orthonormal basis, $\{|u_i\rangle\}$, we can form a two-particle orthonormal basis, $\{|u_i(1); u_j(2)\rangle\}$, where the arguments "1" and "2" refer to the first and second particle. Confining ourselves to a two-particle system, as we do throughout this paper, let $P$ be the Hermitian permutation operator defined by

$$P|u_i(1); u_j(2)\rangle = |u_i(2); u_j(1)\rangle$$

$$= |u_j(1); u_i(2)\rangle. \tag{1}$$

Note that $P$ is involutory, meaning that $P^2 = 1$. A state described by a wave vector $|\Psi\rangle$ is symmetric or antisymmetric if $P|\Psi\rangle = |\Psi\rangle$ or $P|\Psi\rangle = -|\Psi\rangle$, respectively. (A discussion of systems of identical properties is presented, for example, in [1], Chapter XIV.) However, when the state is described by a density operator, we will be more precise and say that $\rho$ is state-symmetric or state-antisymmetric according to whether $P\rho = \rho$ or $P\rho = -\rho$ [12]. An operator $O$ is symmetric if

$$O = POP \tag{2}$$

and antisymmetric if

$$O = -POP, \tag{3}$$

but when condition (2) applies to a density operator, we will use the phrase operator-symmetric to avoid ambiguity. (Any operator that satisfies Eq. (3) is traceless and therefore density operators cannot be operator-antisymmetric.) A density operator can at once be operator-symmetric and state-antisymmetric, for example. We also note that if a density operator is either state-symmetric or antisymmetric then it is operator-symmetric [12]. Something resembling a converse is provided below.

Following standard theory [1], from the one-particle orthonormal basis, $\{|u_i\rangle\}$, we construct two-particle states that have definite symmetry. We assume that we can build a basis from the eigenvectors of the Hermitian operator $P$. The two eigenspaces of $P$ are

$$E_1 = \text{span}\left\{|\Psi_{s;ij}\rangle = \frac{1}{\sqrt{2(1+\delta_{ij})}}(|u_i(1);u_j(2)\rangle + |u_i(2);u_j(1)\rangle), i \leq j\right\} \quad (4)$$

and

$$E_{-1} = \text{span}\left\{|\Psi_{a;ij}\rangle = \frac{1}{\sqrt{2}}(|u_i(1);u_j(2)\rangle - |u_i(2);u_j(1)\rangle), i < j\right\} \quad (5)$$

where $\{|\Psi_{s;ij}\rangle\}$ and $\{|\Psi_{a;ij}\rangle\}$ are orthonormal sets of eigenvectors of $P$ with eigenvalues 1 and $-1$, respectively. (That $i \neq j$ in the eigenspace $E_{-1}$ is a manifestation of the Pauli exclusion principle.) A resolution of the identity is

$$\sum_{i \leq j}|\Psi_{s;ij}\rangle\langle\Psi_{s;ij}| + \sum_{i < j}|\Psi_{a;ij}\rangle\langle\Psi_{a;ij}| = 1 \quad (6)$$

and $P$ itself can be written as

$$P = \sum_{i \leq j}|\Psi_{s;ij}\rangle\langle\Psi_{s;ij}| - \sum_{i < j}|\Psi_{a;ij}\rangle\langle\Psi_{a;ij}|, \quad (7)$$

whence we see that the well-known symmetrizer

$$S = \frac{1}{2}(1+P) \quad (8)$$

and antisymmetrizer

$$A = \frac{1}{2}(1-P) \quad (9)$$

convert an arbitrary two-particle state to a symmetric and antisymmetric one, respectively. (The factor of $1/2$ ensures that the Hermitian operators $S$ and $A$ are projectors, $S = S^2$ and $A = A^2$, and these last two equations also imply neither operator is unitary.) Adding Eqs. (8) and (9) yields

$$A + S = 1, \quad (10)$$

a result that follows for the case under study here (viz., two particles), but which does not generally hold for more than two particles [1].

We can invert bases using

$$|u_i(1); u_j(2)\rangle = \begin{cases} |\Psi_{s;ii}\rangle \text{ if } i = j \\ \frac{1}{\sqrt{2}}(|\Psi_{s;ij}\rangle + |\Psi_{a;ij}\rangle) \text{ if } i < j \\ \frac{1}{\sqrt{2}}(|\Psi_{s;ji}\rangle - |\Psi_{a;ji}\rangle) \text{ if } i > j \end{cases} \quad (11)$$

An arbitrary, two-particle vector $|\Psi\rangle$ can be written as the superposition

$$|\Psi\rangle = \sum_{i,j} c_{ij} |u_i(1); u_j(2)\rangle$$

$$= |\Psi_S\rangle + |\Psi_A\rangle \quad (12)$$

where the symmetric state

$$|\Psi_S\rangle = \sum_{i \leq j} \sqrt{\frac{1}{2(1+\delta_{ij})}} (c_{ij} + c_{ji}) |\Psi_{s;ij}\rangle \quad (13)$$

and the antisymmetric state

$$|\Psi_A\rangle = \frac{1}{\sqrt{2}} \sum_{i<j} (c_{ij} - c_{ji}) |\Psi_{a;ij}\rangle . \quad (14)$$

Since $SA = AS = 0$, and noting that $S|\Psi_S\rangle = |\Psi_S\rangle$ and $A|\Psi_A\rangle = |\Psi_A\rangle$, we confirm that

$$S|\Psi\rangle = |\Psi_S\rangle \quad (15)$$

and

$$A|\Psi\rangle = |\Psi_A\rangle . \quad (16)$$

A somewhat different, but equivalent definition of "state-symmetric" is implicit in the statement that $P\rho = \rho$ is a necessary and sufficient condition for $\frac{1}{2}[P,\rho]_+ = \rho$, given that $\rho = \sum_i p_i |\Phi_i\rangle\langle\Phi_i|$, where $\Sigma_i p_i = 1$, $p_j > 0$ and $\langle\Phi_j|\Phi_j\rangle = 1$ for all $j$. Sufficiency can be seen immediately by recalling that $P$ and $\rho$ are Hermitian so that $\frac{1}{2}P\rho + \frac{1}{2}(P\rho)^\dagger = \frac{1}{2}\rho + \frac{1}{2}\rho^\dagger$ yields the result. Necessity follows from the following steps:

$$\frac{1}{2}[P,\rho]_+ = \rho \quad (17)$$

$\Rightarrow$

$$\text{Tr}(1-P)\rho = 0 \quad (18)$$

$\Rightarrow$

$$\sum_j p_j \langle \Phi_j | A^\dagger A | \Phi_j \rangle = 0 \tag{19}$$

$\Rightarrow$

$$\langle \Phi_j | A^\dagger A | \Phi_j \rangle = 0 \tag{20}$$

$\Rightarrow$

$$|A\Phi_j\rangle = 0 \tag{21}$$

$\Rightarrow$

$$\rho = P\rho. \tag{22}$$

It is not generally true that a density operator $\rho$ is state-symmetric only if it is a mixture of pure density operators each corresponding to a wave vector $|\Psi_{s;ij}\rangle$ (i.e., $\rho \neq \sum_{i \leq j} p_{ij} |\Psi_{s;ij}\rangle\langle\Psi_{s;ij}|$ in general); but it is true that a density operator $\rho = \sum_i p_i |\Phi_i\rangle\langle\Phi_i|$, where $\Sigma_i p_i = 1$, $p_j > 0$ and $\langle\Phi_j|\Phi_j\rangle = 1$ for all $j$, is state-symmetric only if it can be expressed as $\rho = \sum_i p_i |\Phi_{S,i}\rangle\langle\Phi_{S,i}|$ where the $|\Phi_{S,i}\rangle$ are symmetric kets. Recall that the basis $\{u_i\}$ was chosen arbitrarily and therefore we cannot expect $\{|\Psi_{s;ij}\rangle\}$, which was constructed therefrom, to necessarily be the basis that diagonalizes $\rho$. Analogous statements can be made about state-antisymmetric density operators. The foregoing is encapsulated by the following.

Lemma 1: Assume $\rho$ is a density operator and therefore Hermitian, normalized and positive. Then $P\rho = \rho$ is a necessary and sufficient condition for

$$\rho = \sum_{k \leq l, i \leq j} \langle \Psi_{s;ij} | \rho | \Psi_{s;kl} \rangle |\Psi_{s;ij}\rangle\langle\Psi_{s;kl}| \tag{23}$$

where

$$\sum_{i \leq j} \langle \Psi_{s;ij} | \rho | \Psi_{s;ij} \rangle = 1 \tag{24}$$

and

$$\langle \Psi_{s;ij} | \rho | \Psi_{s;kl} \rangle = \langle \Psi_{s;kl} | \rho | \Psi_{s;ij} \rangle^*. \tag{25}$$

Proof:

*sufficiency*

Using the resolution of the identity twice, any density operator can be written as

$$\rho = \sum_{i\leq j}\sum_{k\leq l}\langle\Psi_{s;ij}|\rho|\Psi_{s;kl}\rangle|\Psi_{s;ij}\rangle\langle\Psi_{s;kl}| + \sum_{i<j}\sum_{k\leq l}\langle\Psi_{a;ij}|\rho|\Psi_{s;kl}\rangle|\Psi_{a;ij}\rangle\langle\Psi_{s;kl}|$$
$$+ \sum_{i\leq j}\sum_{k<l}\langle\Psi_{s;ij}|\rho|\Psi_{a;kl}\rangle|\Psi_{s;ij}\rangle\langle\Psi_{a;kl}| + \sum_{i<j}\sum_{k<l}\langle\Psi_{a;ij}|\rho|\Psi_{a;kl}\rangle|\Psi_{a;ij}\rangle\langle\Psi_{a;kl}| \quad (26)$$

We are assuming that $\rho - P\rho = 0$, which implies

$$\sum_{i<j}\sum_{k\leq l}\langle\Psi_{a;ij}|\rho|\Psi_{s;kl}\rangle|\Psi_{a;ij}\rangle\langle\Psi_{s;kl}| + \sum_{i<j}\sum_{k<l}\langle\Psi_{a;ij}|\rho|\Psi_{a;kl}\rangle|\Psi_{a;ij}\rangle\langle\Psi_{a;kl}| = 0 \quad (27)$$

Therefore,

$$\rho = \sum_{i\leq j}\sum_{k\leq l}\langle\Psi_{s;ij}|\rho|\Psi_{s;kl}\rangle|\Psi_{s;ij}\rangle\langle\Psi_{s;kl}| + \sum_{i\leq j}\sum_{k<l}\langle\Psi_{s;ij}|\rho|\Psi_{a;kl}\rangle|\Psi_{s;ij}\rangle\langle\Psi_{a;kl}| \quad (28)$$

This last equation and $\rho = \rho^\dagger$ implies

$$\sum_{i\leq j}\sum_{k<l}\langle\Psi_{s;ij}|\rho|\Psi_{a;kl}\rangle|\Psi_{s;ij}\rangle\langle\Psi_{a;kl}| - \sum_{i\leq j}\sum_{k<l}\langle\Psi_{a;kl}|\rho|\Psi_{s;ij}\rangle|\Psi_{a;kl}\rangle\langle\Psi_{s;ij}| = 0 \quad (29)$$

Letting $S$ act on both sides of this last equation from the left, yields

$$\sum_{i\leq j}\sum_{k<l}\langle\Psi_{s;ij}|\rho|\Psi_{a;kl}\rangle|\Psi_{s;ij}\rangle\langle\Psi_{a;kl}| = 0 \quad (30)$$

Hence,

$$\rho = \sum_{i\leq j}\sum_{k\leq l}\langle\Psi_{s;ij}|\rho|\Psi_{s;kl}\rangle|\Psi_{s;ij}\rangle\langle\Psi_{s;kl}| \quad (31)$$

Normalization and Hermiticity of $\rho$ respectively lead to $\sum_{i\leq j}\langle\Psi_{s;ij}|\rho|\Psi_{s;ij}\rangle = 1$ and $\langle\Psi_{s;ij}|\rho|\Psi_{s;kl}\rangle = \langle\Psi_{s;kl}|\rho|\Psi_{s;ij}\rangle^*$.

*necessity*

Since $|\Psi_{s;ij}\rangle$ is symmetric, $P|\Psi_{s;ij}\rangle = |\Psi_{s;ij}\rangle$. Consequently, if $\rho = \sum_{k\leq l, i\leq j}\langle\Psi_{s;ij}|\rho|\Psi_{s;kl}\rangle|\Psi_{s;ij}\rangle\langle\Psi_{s;kl}|$, then $P\rho = \rho$.

QED

The following lemma, which involves the symmetricity, $\text{Tr}P\rho$, will be useful.

Lemma 2: For a normalized density operator $\rho = \sum_i p_i|\Phi_i\rangle\langle\Phi_i|$, where $\Sigma_i p_i = 1$, $p_j > 0$ and $\langle\Phi_j|\Phi_j\rangle = 1$ for all $j$, we have

$$|\text{Tr}P\rho| \leq 1, \quad (32)$$

where $\text{Tr}P\rho$ is equal to $-1$ if, and only if, $\rho$ is state-antisymmetric and equal to 1 if, and only if, $\rho$ is state-symmetric.

Proof:

First, we show that $\text{Tr}P\rho = 1 \Leftrightarrow P\rho = \rho$.

Trivially, taking the trace of both sides,

$P\rho = \rho \Rightarrow \text{Tr}P\rho = 1$

For the converse, we have to show that $\text{Tr}P\rho = 1 \Rightarrow P\rho = \rho$. We start with $\text{Tr}P\rho = 1$, which implies

$$\sum_j p_j \langle \Phi_j | P | \Phi_j \rangle = \sum_j p_j \langle \Phi_j | \Phi_j \rangle \tag{33}$$

$\Rightarrow$

$$\sum_j p_j \langle \Phi_j | A^\dagger A | \Phi_j \rangle = 0 \tag{34}$$

$\Rightarrow$

$$|A\Phi_j\rangle = 0 \text{, for all } |\Phi_j\rangle \tag{35}$$

$\Rightarrow$

$$A\rho = 0 \tag{36}$$

$\Rightarrow$

$$(A + S)\rho = S\rho \tag{37}$$

$\Rightarrow$

$$\rho = S\rho \tag{38}$$

$\Rightarrow$

$$\rho = P\rho. \tag{39}$$

A similar proof shows that $\text{Tr}P\rho = -1 \Leftrightarrow P\rho = -\rho$. Finally, we have to show that $-1 \leq \text{Tr}P\rho \leq 1$.

Using Eq. (26), we can compute

$$\mathrm{Tr} P\rho = \sum_{i \leq j} \langle \Psi_{s;ij} | \rho | \Psi_{s;ij} \rangle - \sum_{i<j} \langle \Psi_{a;ij} | \rho | \Psi_{a;ij} \rangle \tag{40}$$

We also have

$$\begin{aligned} 1 &= \mathrm{Tr}\rho \\ &= \sum_{i \leq j} \langle \Psi_{s;ij} | \rho | \Psi_{s;ij} \rangle + \sum_{i<j} \langle \Psi_{a;ij} | \rho | \Psi_{a;ij} \rangle \end{aligned} \tag{41}$$

We now make use of the elementary fact that if two non-negative real numbers $a$ and $b$ add to one, their difference lies in the range $-1$ to $1$ since $a + b = 1 \Rightarrow a, b \in [0,1]$ and hence, $a - b = 2a - 1 \in [-1,1]$. Associating $a$ with $\sum_{i \leq j} \langle \Psi_{s,ij} | \rho | \Psi_{s,ij} \rangle$ and $b$ with $\sum_{i<j} \langle \Psi_{a,ij} | \rho | \Psi_{a,ij} \rangle$ yields $-1 \leq \mathrm{Tr} P\rho \leq 1$.

QED

The symmetricity measures the degree to which a density operator is state-symmetric or antisymmetric. A state is perfectly asymmetric (i.e., has no bias towards being symmetric or antisymmetric) if the symmetricity is zero and becomes more symmetric or antisymmetric the closer the symmetricity approaches 1 or $-1$, respectively. This concept is reminiscent of anyonic particles with spin statistics "interpolating continuously between the usual boson and fermion cases." [18] See also [10]. In some circumstances, the appropriate initial condition to take for a state whose symmetric or antisymmetric nature is equally in doubt is one which has a symmetricity of zero; we will encounter such an example in Section V, where we examine the elastic collision of identical particles with an unknown spin state. It should be noted, however, that anyons, and more generally parastatistics, are usually associated with exact (or strong) symmetry, whereas the notion of a perfectly asymmetric density operator is weaker, associated with average (or weak) symmetry [19], [20].

Turning to operators, we first note that there is no unique method to transform an operator $O$ into a symmetric one, $O_S$. For example, both $SOS$ and $AOA$ satisfy Eq. (2). We can arrive at a transform $O \mapsto \mathcal{T}_S(O) = O_S$, where $\mathcal{T}_S$ is the operator symmetrizer and $O_S$ is a symmetric operator, if we insist on some reasonable restrictions. We assume that the one-particle operator $H(1)$ transforms according to

$$\mathcal{T}_S[H(1) \otimes 1(2)] = \lambda[H(1) \otimes 1(2) + 1(1) \otimes H(2)], \tag{42}$$

where $\lambda$ is an arbitrary real constant that we will fix later on physical grounds. If $\{E_i(1); E_j(2)\}$ is a two-particle basis obtained from the eigenvectors $|E_i\rangle$ and eigenvalues $E_i$ of the operator $H(1)$, we allow the operators on either side of Eq. (42) to act on these basis members. By duplicating some of the steps leading to Eq. B-32, p. 1381 of [1], we may then proceed as follows.

$$\mathcal{T}_S[H(1) \otimes 1(2)]|E_i(1); E_j(2)\rangle = \lambda[H(1) \otimes 1(2) + 1(1) \otimes H(2)]|E_i(1); E_j(2)\rangle$$

$$= \lambda(E_i + E_j)|E_i(1); E_j(2)\rangle$$

$$= \lambda[H(1) \otimes 1(2) + P(H(1) \otimes 1(2))P]|E_i(1); E_j(2)\rangle, \quad (43)$$

from which we conclude that

$$\mathcal{T}_S[H(1) \otimes 1(2)] = \lambda[H(1) \otimes 1(2) + P(H(1) \otimes 1(2))P]. \quad (44)$$

To ensure that $\mathcal{T}_S$ preserves normalization of density operators and that $\mathcal{T}_S = \mathcal{T}_S^2$, we take $\lambda = 1/2$. If it is the same operator that transforms $H(1) \otimes 1(2)$ and an arbitrary two-particle $O(1,2)$ operator to a symmetric one, we arrive at

$$\mathcal{T}_S[O(1,2)] = \frac{1}{2}[O(1,2) + PO(1,2)P] \quad (45)$$

However, we remark that in the same way that $|S\psi\rangle$ may not be a physical ket until it is normalized, $\mathcal{T}_S[O(1,2)]$ may not be an appropriate physical observable. For example, if $H(1)$ is a one particle Hamiltonian, then the corresponding physical Hamiltonian for two non-interacting identical particles is $2\mathcal{T}_S[H(1) \otimes 1(2)] = H(1) \otimes 1(2) + H(2) \otimes 1(1)$.

Likewise, although $AOS$ and $SOA$ both satisfy Eq. (3), we settle on

$$\mathcal{T}_A = \frac{1}{2}(1 - P \cdot P) \quad (46)$$

for the operator antisymmetrizer. We note the following identities

$$\mathcal{T}_S(\cdot) = A \cdot A + S \cdot S \quad (47)$$

and

$$\mathcal{T}_A(\cdot) = A \cdot S + S \cdot A, \quad (48)$$

whence $\mathcal{T}_S \mathcal{T}_A = \mathcal{T}_A \mathcal{T}_S = 0$. In summary, the operator symmetrizer and antisymmetrizer given above satisfy

$$\mathcal{T}_S(O) = P\mathcal{T}_S(O)P \quad (49)$$

and

$$\mathcal{T}_A(O) = -P\mathcal{T}_A(O)P, \quad (50)$$

respectively. That the operators $\mathcal{T}_S(O)$, $SOS$ and $AOA$ all satisfy Eq. (2) suggests that there is some freedom in choosing to work with the operator symmetrizer, $\mathcal{T}_S$, or the state vector symmetrizer or antisymmetrizer, $S$ or $A$, when calculating probabilities. We will avail ourselves of this freedom later.

The following lemma characterizes operator-symmetric density operators as statistical mixtures.

Lemma 3: If a density operator $\rho = \Sigma_i p_i |\Psi_i\rangle\langle\Psi_i|$, where $\Sigma_i p_i = 1$, $p_j > 0$ and $\langle\Psi_j|\Psi_j\rangle = 1$ for all $j$, is operator-symmetric, then it is a statistical mixture of state-symmetric and antisymmetric density operators.

Proof:

Expanding on the $\{|u_i(1); u_j(2)\rangle\}$ basis,

$$|\Psi_r\rangle = \sum_{i,j} c_{ij}^{(r)} |u_i(1); u_j(2)\rangle$$

$$= \sum_{i \leq j} \sqrt{\frac{1}{2(1+\delta_{ij})}} \left(c_{ij}^{(r)} + c_{ji}^{(r)}\right) |\Psi_{s;ij}\rangle + \frac{1}{\sqrt{2}} \sum_{i<j} \left(c_{ij}^{(r)} - c_{ji}^{(r)}\right) |\Psi_{a;ij}\rangle$$

$$\equiv |\Psi_{S,r}\rangle + |\Psi_{A,r}\rangle, \tag{51}$$

where, using $\langle\Psi_r|\Psi_r\rangle = 1$, we have

$$\langle\Psi_{S,r}|\Psi_{S,r}\rangle + \langle\Psi_{A,r}|\Psi_{A,r}\rangle = 1. \tag{52}$$

Hence,

$$\rho = \sum_r p_r (|\Psi_{S,r}\rangle + |\Psi_{A,r}\rangle)(\langle\Psi_{S,r}| + \langle\Psi_{A,r}|)$$

$$= \sum_r p_r |\Psi_{S,r}\rangle\langle\Psi_{S,r}| + \sum_r p_r |\Psi_{S,r}\rangle\langle\Psi_{A,r}| + \sum_r p_r |\Psi_{A,r}\rangle\langle\Psi_{S,r}| + \sum_r p_r |\Psi_{A,r}\rangle\langle\Psi_{A,r}|. \tag{53}$$

By hypothesis, the density operator is operator-symmetric, $\rho = P \rho P$

or

$$\sum_r p_r |\Psi_{S,r}\rangle\langle\Psi_{S,r}| + \sum_r p_r |\Psi_{S,r}\rangle\langle\Psi_{A,r}| + \sum_r p_r |\Psi_{A,r}\rangle\langle\Psi_{S,r}| + \sum_r p_r |\Psi_{A,r}\rangle\langle\Psi_{A,r}|$$

$$=$$

$$\sum_r p_r |\Psi_{S,r}\rangle\langle\Psi_{S,r}| - \sum_r p_r |\Psi_{S,r}\rangle\langle\Psi_{A,r}| - \sum_r p_r |\Psi_{A,r}\rangle\langle\Psi_{S,r}| + \sum_r p_r |\Psi_{A,r}\rangle\langle\Psi_{A,r}|, \tag{54}$$

which after canceling terms becomes

$$\sum_r p_r |\Psi_{S,r}\rangle\langle\Psi_{A,r}| + \sum_r p_r |\Psi_{A,r}\rangle\langle\Psi_{S,r}| = 0. \tag{55}$$

Hence,

$$\rho = \sum_r p_r |\Psi_{S,r}\rangle\langle\Psi_{S,r}| + \sum_r p_r |\Psi_{A,r}\rangle\langle\Psi_{A,r}|. \tag{56}$$

For each $r$, at least one of $|\Psi_{A,r}\rangle$ and $|\Psi_{S,r}\rangle$ is non-zero on account of Eq. (52). There are then four cases to consider: a zero ket is found among the elements of a) both the antisymmetric set $\{|\Psi_{A,r}\rangle\}$ and the symmetric set $\{|\Psi_{S,r}\rangle\}$, b) only the antisymmetric set, c) only the symmetric set, and d) neither the antisymmetric nor the symmetric sets.

For case a), we may partition the set of positive integers into three non-ovelapping sets $\{i_1, i_2, \ldots\} \cup \{j_1, j_2, \ldots\} \cup \{k_1, k_2, \ldots\} = \{1, 2, \ldots\}$ according to

$$\langle\Psi_{A,r}|\Psi_{A,r}\rangle = 0 \Rightarrow \langle\Psi_{S,r}|\Psi_{S,r}\rangle = 1 \text{ if } r \in \{i_1, i_2, \ldots\}$$

$$\langle\Psi_{S,r}|\Psi_{S,r}\rangle = 0 \Rightarrow \langle\Psi_{A,r}|\Psi_{A,r}\rangle = 1 \text{ if } r \in \{j_1, j_2, \ldots\}$$

and

$$\langle\Psi_{A,r}|\Psi_{A,r}\rangle\langle\Psi_{S,r}|\Psi_{S,r}\rangle \neq 0 \text{ if } r \in \{k_1, k_2, \ldots\}$$

where we have used Eq. (52). In view of Eq. (56), we can write the density operator as

$$\rho = \sum_{\{i_1,i_2,\ldots\}} p_r |\Psi_{S,r}\rangle\langle\Psi_{S,r}| + \sum_{\{j_1,j_2,\ldots\}} p_r |\Psi_{A,r}\rangle\langle\Psi_{A,r}| + \sum_{\{k_1,k_2,\ldots\}} p_r (|\Psi_{S,r}\rangle\langle\Psi_{S,r}| + |\Psi_{A,r}\rangle\langle\Psi_{A,r}|) \tag{57}$$

Note that

$$\sum_{r=1}^{\infty} p_r = \sum_{\{i_1,i_2,\ldots\}} p_r + \sum_{\{j_1,j_2,\ldots\}} p_r + \sum_{\{k_1,k_2,\ldots\}} p_r = 1 \tag{58}$$

Similarly, we may also write for case b)

$$\rho = \sum_{\{i'_1,i'_2,\ldots\}} p_r |\Psi_{S,r}\rangle\langle\Psi_{S,r}| + \sum_{\{k'_1,k'_2,\ldots\}} p_r (|\Psi_{S,r}\rangle\langle\Psi_{S,r}| + |\Psi_{A,r}\rangle\langle\Psi_{A,r}|) \tag{59}$$

where $\langle\Psi_{A,r}|\Psi_{A,r}\rangle = 0$ if $r \in \{i'_1, i'_2, \ldots\}$ and $\langle\Psi_{A,r}|\Psi_{A,r}\rangle\langle\Psi_{S,r}|\Psi_{S,r}\rangle \neq 0$ if $r \in \{k'_1, k'_2, \ldots\}$,

for case c),

$$\rho = \sum_{\{j''_1,j''_2,\ldots\}} p_r |\Psi_{A,r}\rangle\langle\Psi_{A,r}| + \sum_{\{k''_1,k''_2,\ldots\}} p_r (|\Psi_{S,r}\rangle\langle\Psi_{S,r}| + |\Psi_{A,r}\rangle\langle\Psi_{A,r}|) \tag{60}$$

where $\langle\Psi_{S,r}|\Psi_{S,r}\rangle = 0$ if $r \in \{j''_1, j''_2, \ldots\}$ and $\langle\Psi_{A,r}|\Psi_{A,r}\rangle\langle\Psi_{S,r}|\Psi_{S,r}\rangle \neq 0$ if $r \in \{k''_1, k''_2, \ldots\}$

and for case d),

$$\rho = \sum_{r=1} p_r |\Psi_{S,r}\rangle\langle\Psi_{S,r}| + \sum_{r=1} p_r |\Psi_{A,r}\rangle\langle\Psi_{A,r}| \tag{61}$$

where $\langle\Psi_{A,r}|\Psi_{A,r}\rangle\langle\Psi_{S,r}|\Psi_{S,r}\rangle \neq 0$ for all $r$.

For brevity of the presentation, we will assume that the generic case a) applies, corresponding to Eq. (57), and note that the proof carries, *mutatis mutandis*, for the other cases.

Since $\langle\Psi_{A,r}|\Psi_{A,r}\rangle + \langle\Psi_{S,r}|\Psi_{S,r}\rangle = 1$ and in view of Eq. (57),

$$\rho = \sum_{\{i_1,i_2,\ldots\}} p_r |\Psi_{S,r}\rangle\langle\Psi_{S,r}| + \sum_{\{j_1,j_2,\ldots\}} p_r |\Psi_{A,r}\rangle\langle\Psi_{A,r}|$$

$$+ \sum_{\{k_1,k_2,\ldots\}} \left( \frac{p_r}{\frac{\langle\Psi_{A,r}|\Psi_{A,r}\rangle}{\langle\Psi_{S,r}|\Psi_{S,r}\rangle}+1} \frac{|\Psi_{S,r}\rangle\langle\Psi_{S,r}|}{\langle\Psi_{S,r}|\Psi_{S,r}\rangle} + \frac{p_r}{1+\frac{\langle\Psi_{S,r}|\Psi_{S,r}\rangle}{\langle\Psi_{A,r}|\Psi_{A,r}\rangle}} \frac{|\Psi_{A,r}\rangle\langle\Psi_{A,r}|}{\langle\Psi_{A,r}|\Psi_{A,r}\rangle} \right)$$

$$= \sum_{\{i_1,i_2,\ldots\}} p_r |\Psi_{S,r}\rangle\langle\Psi_{S,r}| + \sum_{\{j_1,j_2,\ldots\}} p_r |\Psi_{A,r}\rangle\langle\Psi_{A,r}| + \sum_{\{k_1,k_2,\ldots\}} p_{r,S} \frac{|\Psi_{S,r}\rangle\langle\Psi_{S,r}|}{\langle\Psi_{S,r}|\Psi_{S,r}\rangle} + \sum_{\{k_1,k_2,\ldots\}} p_{r,A} \frac{|\Psi_{A,r}\rangle\langle\Psi_{A,r}|}{\langle\Psi_{A,r}|\Psi_{A,r}\rangle} \tag{62}$$

where

$$0 < p_{r,S} = \frac{p_r}{\frac{\langle\Psi_{A,r}|\Psi_{A,r}\rangle}{\langle\Psi_{S,r}|\Psi_{S,r}\rangle}+1} < 1$$

and

$$0 < p_{r,A} = \frac{p_r}{1+\frac{\langle\Psi_{S,r}|\Psi_{S,r}\rangle}{\langle\Psi_{A,r}|\Psi_{A,r}\rangle}} < 1$$

for $r \in \{k_1, k_2, \ldots\}$.

With the last two equations and Eq. (58) we may confirm that

$$\sum_{\{i_1,i_2,\ldots\}} p_r + \sum_{\{j_1,j_2,\ldots\}} p_r + \sum_{\{k_1,k_2,\ldots\}} p_{r,S} + \sum_{\{k_1,k_2,\ldots\}} p_{r,A} = 1 \tag{63}$$

The right-hand side of Eq. (62) is a statistical mixture of Hermitian, positive and normalized operators---i.e., density operators---that are state-symmetric or antisymmetric.

QED

With symmetric and antisymmetric states $|\Phi_S\rangle = S|\Phi\rangle$ and $|\Phi_A\rangle = A|\Phi\rangle$ (and similar expressions replacing $\Phi$ by $\Psi$), and symmetric and antisymmetric operators, $O_S = \frac{1}{2}(O + POP)$ and $O_A = \frac{1}{2}(O - POP)$, we may compute matrix elements in a Schrodinger or Heisenberg picture. To wit,

$$\langle\Psi_S|O|\Phi_S\rangle = \tfrac{1}{2}\langle\Psi|\left(O_S + \tfrac{1}{2}[P, O_S]_+\right)|\Phi\rangle \tag{64}$$

$$\langle\Psi_A|O|\Phi_A\rangle = \tfrac{1}{2}\langle\Psi|\left(O_S - \tfrac{1}{2}[P, O_S]_+\right)|\Phi\rangle \tag{65}$$

$$\langle\Psi_S|O|\Phi_A\rangle = \tfrac{1}{2}\langle\Psi|\left(O_A + \tfrac{1}{2}[P, O_A]\right)|\Phi\rangle \tag{66}$$

$$\langle\Psi_A|O|\Phi_S\rangle = \tfrac{1}{2}\langle\Psi|\left(O_A - \tfrac{1}{2}[P, O_A]\right)|\Phi\rangle \tag{67}$$

The matrix elements on the left-hand side of these equations are in the Schrodinger picture whereas those on the right-hand side are in the Heisenberg picture since the symmetric structure is carried in the state vectors and operators, respectively. Traditional formulations use symmetric operators in a mixed picture, focusing on $\langle\Psi_S|O_S|\Phi_S\rangle$ $(= \langle\Psi_S|O|\Phi_S\rangle)$ and $\langle\Psi_A|O_S|\Phi_A\rangle$ $(= \langle\Psi_A|O|\Phi_A\rangle)$.

### III. **Symmetrization of Operators**

In this section, we examine whether a completely positive dynamical map, arising from environmental coupling, can symmetrize or antisymmetrize an operator. We also investigate whether such a map can have the semigroup property.

Our starting point is the following identity:

$$e^{\tau P\cdot P}\rho = \rho + \tfrac{\tau}{1!}P\rho P + \tfrac{\tau^2}{2!}\rho + \tfrac{\tau^3}{3!}P\rho P + \ldots$$

$$= \cosh(\tau)\rho + \sinh(\tau)P\rho P, \tag{68}$$

whence

$$e^{-\tau}e^{\tau P\cdot P}\rho \sim \tfrac{1}{2}(1 + P\cdot P)\rho, \text{ as } \tau \to \infty, \tag{69}$$

which is the operator symmetrizer.

Using notation that we have used elsewhere [21],

$$\{A, \rho, B\} = BA^\dagger\rho + \rho BA^\dagger - 2A^\dagger\rho B, \tag{70}$$

we note that

$$\{P, \rho, P\} = 2\rho - 2P\rho P \tag{71}$$

and hence

$$e^{-\frac{\tau}{2}\{P,\cdot,P\}} = e^{-\tau} e^{\tau(P \cdot P)} \tag{72}$$

We conclude that $e^{-\frac{\tau}{2}\{P,\cdot,P\}}$ is an asymptotic expression for the operator symmetrizer,

$$e^{-\frac{\tau}{2}\{P,\cdot,P\}}(\rho) \sim \mathcal{T}_S(\rho), \text{ as } \tau \to \infty \tag{73}$$

$$= \frac{1}{2}(\rho + P\rho P).$$

This last expression is in the form of a positive operator-sum, and the left-hand side of relation (73) may also be written in this form [22], [21], which then leads to

$$\frac{1}{\sqrt{2\pi\tau}} \int_{-\infty}^{\infty} du \exp\left(-\frac{u^2}{2\tau}\right) e^{-iPu} \rho e^{iPu} \sim \frac{1}{2}(\rho + P\rho P), \text{ as } \tau \to \infty. \tag{74}$$

In a similar fashion, we may develop an asymptotic expression for the operator antisymmetrizer,

$$e^{-2\tau} e^{\frac{\tau}{2}\{P,\cdot,P\}}(O) \sim \frac{1}{2}(O - POP) \text{ as } \tau \to \infty. \tag{75}$$

We note that the operator $e^{\frac{\tau}{2}\{P,\cdot,P\}}$ is the inverse of $e^{-\frac{\tau}{2}\{P,\cdot,P\}}$, although it is not defined on the set of all density operators. We may write the last asymptotic expression as [21]

$$\frac{1}{\sqrt{2\pi\tau}} e^{-2\tau} \int_{-\infty}^{\infty} du \exp\left(-\frac{u^2}{2\tau}\right) e^{uP} O e^{-uP} \sim \frac{1}{2}(O - POP) \text{ as } \tau \to \infty; \tag{76}$$

however, this is not a positive operator-sum since $e^{uP}$ and $e^{-uP}$ are not adjoints of each other, resulting in the last integral not being generally positivity-preserving. These observations are consistent with the fact that the operator symmetrizer is a manifestly positive map, but the antisymmetrizer is not.

The asymptotic approximation for the operator symmetrizer, $e^{-\frac{\tau}{2}\{P,\cdot,P\}}$, leads to decoherence of density operators. With the help of Eq. (74), we can calculate $e^{-\frac{\tau}{2}\{P,\cdot,P\}} \rho$ for a general density operator given by Eq. (26):

$$e^{-\frac{\tau}{2}\{P,\cdot,P\}} \rho = \sum_{i \leq j} \sum_{k \leq l} \langle \Psi_{s;ij} | \rho | \Psi_{s;kl} \rangle |\Psi_{s;ij}\rangle\langle\Psi_{s;kl}| + \sum_{i < j} \sum_{k < l} \langle \Psi_{a;ij} | \rho | \Psi_{a;kl} \rangle |\Psi_{a;ij}\rangle\langle\Psi_{a;kl}|$$

$$+ e^{-2\tau} \left( \sum_{i < j} \sum_{k \leq l} \langle \Psi_{a;ij} | \rho | \Psi_{s;kl} \rangle |\Psi_{a;ij}\rangle\langle\Psi_{s;kl}| + \sum_{i \leq j} \sum_{k < l} \langle \Psi_{s;ij} | \rho | \Psi_{a;kl} \rangle |\Psi_{s;ij}\rangle\langle\Psi_{a;kl}| \right). \tag{77}$$

Letting $|\Phi_S\rangle$ and $|\Psi_A\rangle$ be a symmetric and antisymmetric state respectively, we obtain the following matrix element,

$$\langle\Psi_A|e^{-\frac{\tau}{2}\{P,\cdot,P\}}\rho|\Phi_S\rangle = \langle\Psi_A|\rho|\Phi_S\rangle e^{-2\tau}. \tag{78}$$

Off-diagonal elements involving inner products of mixed symmetry decay exponentially. As provided below, $\tau$ may be expressed in terms of the parameters of a microscopic model.

To this end, we now show that the operator symmetrizer arises from the interaction between two particles and their environment. To begin, we introduce the following total Hamiltonian,

$$H_T = H + H_R + V_I + \sum_k \frac{c_k^2}{\hbar\omega_k}. \tag{79}$$

where $H$ is a symmetric two-particle Hamiltonian ($[H,P]=0$), $H_R = \sum_k \hbar\omega_k b_k^\dagger b_k$ is a reservoir Hamiltonian modelling the environment (the operator $b_k$ is the lowering operator of the $k$th mode), and $V_I = P\sum_k c_k(b_k^\dagger + b_k)$ is an interaction term with $c_k$ real. The last "renormalization" term in Eq. (79) will cancel in the following steps [23]. Because $[H, V_I] = 0$, we have what is known in the literature as a quantum non-demolition (QND) coupling. This will allow us to follow the general analysis of this type of coupling found in Ref. [23], from which the ensuing QND work obtains.

We are interested in acquiring the reduced density operator $\rho(t) = \text{Tr}_R \rho_T(t)$ where $\rho_T(t)$ is the density operator of the total system evolving under $H_T$ and the trace is over reservoir variables. The unitary operator

$$\widetilde{U}_T = \exp\left[P\sum_k \frac{c_k}{\hbar\omega_k}(b_k^\dagger - b_k)\right] \tag{80}$$

is useful for this purpose because it decouples the total Hamiltonian according to [23]

$$\widetilde{U}_T H_T \widetilde{U}_T^{-1} = H + H_R \tag{81}$$

$$\equiv \widetilde{H}_T.$$

Denoting the interaction reduced density operator

$$\rho^I(t) = e^{iHt/\hbar}\rho(t)e^{-iHt/\hbar}, \tag{82}$$

we may compute the matrix elements

$$\langle\Psi_{s,ij}|\rho^I(t)|\Psi_{s,kl}\rangle_a = \langle\Psi_{s,ij}|\rho^I(0)|\Psi_{s,kl}\rangle_a Tr_R \widetilde{U}_\leftarrow^{-1} e^{-iH_Rt/\hbar}\widetilde{U}_\leftarrow \rho_R \widetilde{U}_\rightarrow^{-1} e^{iH_Rt/\hbar}\widetilde{U}_\rightarrow \quad (83)$$

with

$$\widetilde{U}_\leftarrow = \exp\left[\pm\sum_l \frac{c_l}{\hbar\omega_l}(b_l^\dagger - b_l)\right], \quad (84)$$

where the top sign (+) goes with the bra $\langle\Psi_{s,ij}|$ and the bottom sign (−) goes with the bra $\langle\Psi_{a,ij}|$ (the arrow points left for the bra),

$$\widetilde{U}_\leftarrow^{-1} = \left(\widetilde{U}_\leftarrow\right)^{-1}, \quad (85)$$

$$\widetilde{U}_\rightarrow = \exp\left[\pm\sum_j \frac{c_j}{\hbar\omega_j}(b_j^\dagger - b_j)\right], \quad (86)$$

where the top sign goes with the ket $|\Psi_{s,kl}\rangle$ and the bottom sign goes with the ket $|\Psi_{a,kl}\rangle$ (the arrow points right for the ket), and

$$\widetilde{U}_\rightarrow^{-1} = \left(\widetilde{U}_\rightarrow\right)^{-1}. \quad (87)$$

Inserting $e^{-iH_Rt/\hbar}e^{iH_Rt/\hbar}$ on the right of the left-hand side of the following equation and using Glauber's formula [1], we get

$$\widetilde{U}_\rightarrow^{-1} e^{iH_Rt/\hbar}\widetilde{U}_\rightarrow =$$

$$\prod_k \left\{\exp\left[\left(\frac{c_k}{\hbar\omega_k}\right)^2 (e^{i\omega_k t} - 1)\right] \exp\left[\pm\frac{c_k}{\hbar\omega_k}(e^{i\omega_k t} - 1)b_k^\dagger\right] \exp\left[\pm\frac{c_k}{\hbar\omega_k}(1 - e^{-i\omega_k t})b_k\right] e^{i\hbar\omega_k b_k^\dagger b_k t/\hbar}\right\} \quad (88)$$

with the same expression for $\widetilde{U}_\leftarrow^{-1} e^{-iH_Rt/\hbar}\widetilde{U}_\leftarrow$. This allows us to write the matrix element as

$$\langle\Psi_{s,ij}|\rho^I(t)|\Psi_{s,kl}\rangle_a =$$

$$\langle\Psi_{s,ij}|\rho^I(0)|\Psi_{s,kl}\rangle_a \prod_k Tr_k \left\{\begin{array}{c} \exp\left[2[(\leftarrow\pm)(\pm\rightarrow) - 1]\left(\frac{c_k}{\hbar\omega_k}\right)^2(1 - \cos\omega_k t)\right] \\ \exp\left[[(\leftarrow\pm) + (\mp\rightarrow)]\frac{c_k}{\hbar\omega_k}(1 - e^{i\omega_k t})b_k^\dagger\right] \\ \exp\left[[(\leftarrow\mp) + (\pm\rightarrow)]\frac{c_k}{\hbar\omega_k}(1 - e^{-i\omega_k t})b_k\right]\frac{e^{-\beta\hbar\omega_k b_k^\dagger b_k}}{Tr_k e^{-\beta\hbar\omega_k b_k^\dagger b_k}} \end{array}\right\} \quad (89)$$

We may now use the identity [23]

$$Tr_k\left[\exp(p_k b_k^\dagger)\exp(q_k b_k)\frac{e^{-\beta\hbar\omega_k b_k^\dagger b_k}}{Tr_k e^{-\beta\hbar\omega_k b_k^\dagger b_k}}\right] = \exp\left(\frac{q_k p_k}{e^{\beta\hbar\omega_k}-1}\right) \quad (90)$$

to take the trace:

$$\left\langle \Psi_{s,ij}^a \left| \rho^I(t) \right| \Psi_{s,kl}^a \right\rangle = \left\langle \Psi_{s,ij}^a \left| \rho^I(0) \right| \Psi_{s,kl}^a \right\rangle \exp\left\{2[(\leftarrow\pm)(\pm\rightarrow)-1]\sum_k \left(\frac{c_k}{\hbar\omega_k}\right)^2(1-\cos\omega_k t)\coth\frac{\beta\hbar\omega_k}{2}\right\} \quad (91)$$

As is usually done at this stage, we approximate the discrete sum over modes by an integral, introducing a density of states in the process. We will utilize the density of states [24], [25]

$$\eta\frac{\omega}{c^2(\omega)}e^{-\omega/\omega_c}, \quad (92)$$

where $\omega_c$ corresponds to a high-frequency cut-off. We compute the integral

$$I(t) \equiv \frac{\eta}{\hbar^2}\int_0^\infty e^{-\omega/\omega_c}\frac{1}{\omega}(1-\cos\omega t)\coth(\beta\hbar\omega/2)\,d\omega \quad (93)$$

or

$$\begin{aligned}\frac{dI}{dt} &= \frac{2\eta}{\hbar^2}\int_0^\infty e^{-\omega/\omega_c}\sin\omega t\frac{1}{e^{\beta\hbar\omega}-1}d\omega + \frac{\eta}{\hbar^2}t\frac{\omega_c^2}{\omega_c^2 t^2+1}\\ &= \frac{2\eta t}{\hbar^2}\sum_{n=1}^\infty \frac{\omega_c^2}{\omega_c^2 t^2+(1+n\beta\hbar\omega_c)^2} + \frac{\eta t}{\hbar^2}\frac{\omega_c^2}{\omega_c^2 t^2+1}\end{aligned} \quad (94)$$

where we have expanded

$\frac{1}{e^{\beta\hbar\omega}-1} = e^{-\beta\hbar\omega}\left(1+e^{-\beta\hbar\omega}+e^{-2\beta\hbar\omega}+e^{-3\beta\hbar\omega}+\ldots\right)$ for $e^{-\beta\hbar\omega} < 1$, due care being given to the fact that the lower limit of integration is $\omega = 0$, which is outside the interval of convergence of the last series.

Integrating, we arrive at

$$I(t) = \frac{\eta}{2\hbar^2}\ln(1+\omega_c^2 t^2) + \frac{\eta}{\hbar^2}\ln\prod_{j=1}^\infty\left(1+\frac{\omega_c^2 t^2}{(j\beta\hbar\omega_c+1)^2}\right), \quad (95)$$

which appears in an equation for $A_{mn}^2$ in [23].

If we now assume that $\beta\hbar\omega_c \gg 1$, we may use an example of the Weierstrass factorization theorem [26],

$$\prod_{j=1}^{\infty}\left[1 + \frac{1}{(\pi j)^2}\left(\frac{\pi t}{\beta\hbar}\right)^2\right] = \frac{\sinh\left(\frac{\pi t}{\beta\hbar}\right)}{\frac{\pi t}{\beta\hbar}}, \tag{96}$$

to obtain (cf. the last equation for $A_{mn}^2$ in [23])

$$I(t) \approx \frac{\eta}{2\hbar^2}\ln(\omega_c^2 t^2 + 1) + \frac{\eta}{\hbar^2}\ln\left(\frac{\beta\hbar}{\pi t}\sinh\frac{\pi t}{\beta\hbar}\right). \tag{97}$$

We can thus approximate

$$\left\langle \Psi_{s,ij \atop a}\left|\rho^I(t)\right|\Psi_{s,kl \atop a}\right\rangle \approx$$

$$\left\langle \Psi_{s,ij \atop a}\left|\rho^I(0)\right|\Psi_{s,kl \atop a}\right\rangle \exp\left\{2[(_\leftarrow\pm)(\pm_\rightarrow) - 1]\left[\frac{\eta}{2\hbar^2}\ln(\omega_c^2 t^2 + 1) + \frac{\eta}{\hbar^2}\ln\frac{\sinh\left(\frac{\pi t}{\beta\hbar}\right)}{\frac{\pi t}{\beta\hbar}}\right]\right\} \tag{98}$$

where $2[(_\leftarrow\pm)(\pm_\rightarrow) - 1] = 0$ for "diagonal" elements (meaning when the bra and the ket are both symmetric or both antisymmetric) and $-4$ for off-diagonal elements. If we assume that

$$\omega_c t \gg \frac{t}{\hbar\beta} \gg 1, \tag{99}$$

which subsumes the condition $\beta\hbar\omega_c \gg 1$ that was used above, we may approximate

$$\left\langle \Psi_{s,ij \atop a}\left|\rho^I(t)\right|\Psi_{s,kl \atop a}\right\rangle \approx \left\langle \Psi_{s,ij \atop a}\left|\rho^I(0)\right|\Psi_{s,kl \atop a}\right\rangle\left(\frac{2\pi}{\beta\hbar\omega_c}\right)^{\frac{2\eta}{\hbar^2}[1-(_\leftarrow\pm)(\pm_\rightarrow)]}\exp\left\{-[1 - (_\leftarrow\pm)(\pm_\rightarrow)]\frac{2\pi\eta t}{\hbar^3\beta}\right\} \tag{100}$$

We note the non-uniformity of this approximation as $t \to 0$, similar to the artefact dubbed a "slip" in Ref. [27]: the conditions (99) make clear that expression (100) is not valid for vanishing time. Finally, to arrive at a semigroup result, we can Taylor expand the following pre-factor in the variable $\frac{2\eta}{\hbar^2}[1 - (_\leftarrow\pm)(\pm_\rightarrow)]\ln\left(\frac{2\pi}{\beta\hbar\omega_c}\right)$ as follows.

$$\left(\frac{2\pi}{\beta\hbar\omega_c}\right)^{\frac{2\eta}{\hbar^2}[1-(_\leftarrow\pm)(\pm_\rightarrow)]} =$$

$$1 + \frac{2\eta}{\hbar^2}[1 - (_\leftarrow\pm)(\pm_\rightarrow)]\ln\left(\frac{2\pi}{\beta\hbar\omega_c}\right) + \frac{1}{2!}\left[\frac{2\eta}{\hbar^2}[1 - (_\leftarrow\pm)(\pm_\rightarrow)]\ln\left(\frac{2\pi}{\beta\hbar\omega_c}\right)\right]^2 + \cdots \tag{101}$$

If we finally further assume that

$$\frac{4\eta}{\hbar^2}\ln\frac{\beta\hbar\omega_c}{2\pi} \ll 1, \tag{102}$$

keeping only the leading term in the series (101), we observe semigroup behaviour (cf. [28] in the context of a QND model),

$$\left\langle \Psi_{s,ij \atop a}\left|\rho^I(t)\right|\Psi_{s,kl \atop a}\right\rangle \approx \left\langle \Psi_{s,ij \atop a}\left|\rho^I(0)\right|\Psi_{s,kl \atop a}\right\rangle \exp\left\{-[1-(_\leftarrow \pm)(\pm_\rightarrow)]\frac{2\pi\eta t}{\hbar^3\beta}\right\} \tag{103}$$

The authors of [23] failed to arrive at the semigroup regime because they did not invoke assumptions, such as all the assumptions (99) and (102), that lead thereto.

Referring to relation (103) we see that $\tau$ in Eq. (78) is equal to $\frac{2\eta\pi t}{\hbar^3\beta}$, and the semigroup master equation that yields relation (103) is

$$\frac{d\rho}{dt} = \frac{1}{\hbar i}[H,\rho] - \frac{\eta\pi}{\hbar^3\beta}\{P,\rho,P\} \tag{104}$$

This last equation preserves the normalization and positivity of the density operator [29].

The qualitative behaviour of the matrix elements varies depending on the assumptions we make involving time, temperature and the high frequency cut-off. However, one unexpected result, whose derivation we omit, arises by taking the different (but still Ohmic) density of states $\eta\frac{\omega}{c^2(\omega)}\theta(\omega_c - \omega)$, with $\theta(\omega)$ being the Heaviside step function: if instead of the foregoing approximations we had assumed $\beta\hbar\omega_c \to 0$, followed by $\omega_c t \to \infty$, we would still have obtained the same Eq. (103). That we recover the same expression for the matrix elements is unanticipated because above we took what in some sense is the opposite approximation, $\beta\hbar\omega_c \gg 1$. Thus, there are at least two disparate paths to the semigroup regime.

## IV.    State Symmetrization via a Completely Positive Dynamical Map

It has been stated that "positivity is not sufficient for a dynamical map to have a physical interpretation for quantum systems, and that we must demand the stronger property of complete positivity." [29] In the previous section, we derived from an interacting environmental model a special type of completely positive dynamical map---a semigroup map---that symmetrizes operators. We now ask whether it is possible for a completely positive dynamical map to symmetrize a state.

To monitor the progress of state symmetrization or antisymmetrization, we can compute the symmetricity as a function of time. To not bias the initial state, we will draw our initial state from the subset of density operators that have zero symmetricity---the perfectly asymmetric states---and investigate the subsequent evolution thereof. For this subset, we will derive an operator-sum expression to state-symmetrize or antisymmetrize density operators.

Suppose we have the conventional [15] completely positive dynamical map

$$\rho(t) = \sum_\alpha \overline{W}_\alpha(t) \rho \overline{W}_\alpha^\dagger(t) \qquad (105)$$

with

$$\sum_\alpha \overline{W}_\alpha^\dagger(t) \overline{W}_\alpha(t) = 1 \qquad (106)$$

and $\rho(0) = \rho$. Then the symmetricity of the density operator is conserved if, for all $\alpha$ and $t \geq 0$, $\overline{W}_\alpha(t)$ is either a symmetric or an antisymmetric operator. To see this, note that if $\overline{W}_\alpha(t)$ is symmetric then $[\overline{W}_\alpha(t), P] = 0$, and thus

$\text{Tr} P \rho(t) = \text{Tr} \sum_\alpha \overline{W}_\alpha^\dagger(t) \overline{W}_\alpha(t) P \rho$

$= \text{Tr} P \rho$

Likewise, if $\overline{W}_\alpha(t)$ is antisymmetric then $[\overline{W}_\alpha(t), P]_+ = 0$ and

$\text{Tr} P \rho(t) = -\text{Tr} P \rho.$

Evaluating the left-hand side at $t = 0$ leads to

$\text{Tr} P \rho = 0$

$\Rightarrow$

$\text{Tr} P \rho(t) = \text{Tr} P \rho$

$= 0.$

It follows that if the symmetricity is to vary, either at least one of the operators $\overline{W}_\alpha(t)$ must have no definite symmetry, or $\sum_\alpha \overline{W}_\alpha^\dagger(t) \overline{W}_\alpha(t) = 1$ must be false at some time. We will give up the latter equation.

From where does Eq. (106) come? Clearly it is sufficient to preserve normalization ($\text{Tr}\rho(t) = 1$), but is it necessary? If we were to insist that $\text{Tr}\rho(t) = \text{Tr}\rho$ for all density operators $\rho$, and assuming that the Hermitian operator $\sum_\alpha \overline{W}_\alpha^\dagger(t)\overline{W}_\alpha(t) - 1 = \sum_k \lambda_k(t)|\lambda_k(t)\rangle\langle\lambda_k(t)|$ for an orthonormal set $\{|\lambda_k(t)\rangle\}$, then we require Eq. (106). But Eq. (106) is not necessary if this restriction is to apply to only a (preferably physically relevant) subset of density operators. In what follows that subset will be initial density operators that are perfectly asymmetric, in other words, that satisfy $\text{Tr}P\rho = 0$.

Before we introduce operators $W_\alpha$ for use in an operator-sum dynamical map, let us pause to examine the class of density operators that are both perfectly asymmetric and operator-symmetric. We can demonstrate that any member of this class, $\rho_{paos}$, is an equally-weighted statistical mixture of a state-antisymmetric and state-symmetric density operator,

$$\rho_{paos} = \frac{1}{2}\rho_A + \frac{1}{2}\rho_S \tag{107}$$

with

$$\rho_A = 2 \sum_{k<l, i<j} \langle \Psi_{a;ij}|\rho_{paos}|\Psi_{a;kl}\rangle |\Psi_{a;ij}\rangle\langle \Psi_{a;kl}| \tag{108}$$

and

$$\rho_S = 2 \sum_{k\leq l, i\leq j} \langle \Psi_{s;ij}|\rho_{paos}|\Psi_{s;kl}\rangle |\Psi_{s;ij}\rangle\langle \Psi_{s;kl}| \tag{109}$$

(cf. Lemma 3).

To wit, from the general expression for a density operator that is Eq. (26), $\rho = P\rho P$ and the linear independence of the set $\{|\Psi_{s,ij}\rangle\langle\Psi_{a,kl}|, |\Psi_{a,i'j'}\rangle\langle\Psi_{s,k'l'}|\}$, we establish that operator-symmetric density operators can be written as

$$\sum_{k<l, i<j} \langle \Psi_{a;ij}|\rho|\Psi_{a;kl}\rangle |\Psi_{a;ij}\rangle\langle \Psi_{a;kl}| + \sum_{k\leq l, i\leq j} \langle \Psi_{s;ij}|\rho|\Psi_{s;kl}\rangle |\Psi_{s;ij}\rangle\langle \Psi_{s;kl}|$$

$$\equiv \frac{1}{2}\rho_A + \frac{1}{2}\rho_S. \tag{110}$$

Again taking an arbitrary $\rho$ given by Eq. (26), the relations $\text{Tr}\rho = 1$ and $\text{Tr}P\rho = 0$ respectively imply

$$\sum_{i\leq j}\langle\Psi_{s;ij}|\rho|\Psi_{s;ij}\rangle + \sum_{i<j}\langle\Psi_{a;ij}|\rho|\Psi_{a;ij}\rangle = 1 \tag{111}$$

and

$$\sum_{i\leq j}\langle\Psi_{s;ij}|\rho|\Psi_{s;ij}\rangle = \sum_{i<j}\langle\Psi_{a;ij}|\rho|\Psi_{a;ij}\rangle \tag{112}$$

from which we find

$$\sum_{i \leq j} \langle \Psi_{s;ij} | \rho | \Psi_{s;ij} \rangle = \sum_{i < j} \langle \Psi_{a;ij} | \rho | \Psi_{a;ij} \rangle = \frac{1}{2} \tag{113}$$

Therefore, $\mathrm{Tr}\rho_A = 1$ and $\mathrm{Tr}\rho_S = 1$. Hermiticity of $\rho_A$ and $\rho_S$ follow from the implied Hermiticity conditions of Eq. (26),

$$\langle \Psi_{a;ij} | \rho | \Psi_{a;kl} \rangle = \langle \Psi_{a;kl} | \rho | \Psi_{a;ij} \rangle^* \tag{114}$$

and

$$\langle \Psi_{s;ij} | \rho | \Psi_{s;kl} \rangle = \langle \Psi_{s;kl} | \rho | \Psi_{s;ij} \rangle^* . \tag{115}$$

Finally, $\rho_A$ is positive if

$\langle \Phi | \rho_A | \Phi \rangle \geq 0$ for all $\Phi$

$\Leftrightarrow$

$$\langle \Phi | A \sum_{i<j} \sum_{k<l} \langle \Psi_{a;ij} | \rho | \Psi_{a;kl} \rangle | \Psi_{a;ij} \rangle \langle \Psi_{a;kl} | A | \Phi \rangle \geq 0 \tag{116}$$

But the assumed positivity of $\rho$ given by Eq. (26) implies

$$\langle \Phi | A \rho A | \Phi \rangle \geq 0$$

$\Rightarrow$

$$\langle \Phi | A \sum_{i<j} \sum_{k<l} \langle \Psi_{a;ij} | \rho | \Psi_{a;kl} \rangle | \Psi_{a;ij} \rangle \langle \Psi_{a;kl} | A | \Phi \rangle \geq 0 \tag{117}$$

which proves $\rho_A$ is positive. A similar argument proves $\rho_S$ is positive. These properties taken together demonstrate that $\rho_A$ and $\rho_S$ are bona fide state-antisymmetric and state-symmetric density operators, respectively.

Returning to our study of a dynamical map that can symmetrize states, let us introduce the following ansatz

$$W_\alpha(t) \equiv a_\alpha(t) A + s_\alpha(t) S, \tag{118}$$

and let $\mathcal{B}_+(\mathcal{H})$ be the set of Hermitian and positive operators on the Hilbert space $\mathcal{H}$ that have finite, though not necessarily unit, trace.

Construct an operator-sum map $\Lambda_t : \mathcal{B}_+(\mathcal{H}) \to \mathcal{B}_+(\mathcal{H})$ using the operators $W_\alpha(t)$ given by Eq. (118),

$$\Lambda_t(\sigma) = \sigma(t)$$

$$= \sum_\alpha (a_\alpha(t)A + s_\alpha(t)S)\sigma(a_\alpha^*(t)A + s_\alpha^*(t)S), \tag{119}$$

where $\sigma \in \mathcal{B}_+(\mathcal{H})$ and the components of the vectors $\mathbf{a} = (a_1, a_2, \ldots)$ and $\mathbf{s} = (s_1, s_2, \ldots)$ are complex numbers to which we will progressively add constraints, which are necessary due to the fact that $\sum_\alpha W_\alpha^\dagger(t) W_\alpha(t)$ is not equal to one for the $W_\alpha(t)$ that we will be interested in.

For continuity, we take

$$\lim_{t \to 0^+} a_\alpha(t) = a_\alpha(0) \tag{120}$$

and

$$\lim_{t \to 0^+} s_\alpha(t) = s_\alpha(0). \tag{121}$$

We also insist that $\sigma(0) = \sigma$ for which it suffices that

$$a_\alpha(0) = s_\alpha(0) \tag{122}$$

and

$$a^2(0) = s^2(0) = 1 \tag{123}$$

where

$$a^2(t) = \|\mathbf{a}(t)\|^2 = \sum_\alpha |a_\alpha(t)|^2 \tag{124}$$

and

$$s^2(t) = \|\mathbf{s}(t)\|^2 = \sum_\alpha |s_\alpha(t)|^2. \tag{125}$$

From Eq. (119), we obtain

$$\mathrm{Tr}\sigma(t) = p(t)\mathrm{Tr}\sigma + m(t)\mathrm{Tr}P\sigma \tag{126}$$

and

$$\mathrm{Tr}P\sigma(t) = m(t)\mathrm{Tr}\sigma + p(t)\mathrm{Tr}P\sigma \tag{127}$$

where

$p(t) \equiv \frac{1}{2}[s^2(t) + a^2(t)]$

and

$m(t) \equiv \frac{1}{2}[s^2(t) - a^2(t)]$.

Condition (123) gives

$p(0) = 1$ and $m(0) = 0$

$\Rightarrow$

$\text{Tr}\sigma(0) = \text{Tr}\sigma$.

If we insist that the trace remain constant, $\text{Tr}\sigma(t) = \text{Tr}\sigma$, we require from Eq. (126),

$$p(t, r_\sigma) = 1 - r_\sigma m(t) \tag{128}$$

where $r_\sigma \equiv \frac{\text{Tr}P\sigma}{\text{Tr}\sigma}$ assuming $\text{Tr}\sigma \neq 0$. One solution of Eq. (128) is $a^2(t) = s^2(t) = 1$. But because we want the symmetricity to vary, we are interested in other solutions that involve $r_\sigma$, and to that end, we have introduced $r_\sigma$ as an argument in $p(t)$ to not lose sight of the dependence thereon. On account of Eq.(128), $a_\alpha$ and $s_\alpha$ will also depend on $t$ and $r_\sigma$, and we likewise write $a_\alpha(t, r_\sigma)$ and $s_\alpha(t, r_\sigma)$ in the following.

In addition to $m(0) = 0$, there is a further constraint on $m(t)$ that arises after inserting Eq. (128) into Eq. (127) and heeding inequality (32):

$$-1 \leq m(t)(1 - r_\sigma^2) + r_\sigma \leq 1 \tag{129}$$

This last expression does not impose any restriction on $m(t)$ when $r_\sigma = \pm 1$. When $|r_\sigma| \neq 1$, relation (129) can be written as

$$-\frac{1}{1 - r_\sigma} \leq m(t) \leq \frac{1}{1 + r_\sigma} \tag{130}$$

Now, consider another non-zero operator $\sigma' \in \mathcal{B}_+(\mathcal{H})$ that we also map according to Eq. (119):

$$\sigma'(t) = \sum_\alpha (a_\alpha(t, r_\sigma)A + s_\alpha(t, r_\sigma)S)\sigma'(a_\alpha^*(t, r_\sigma)A + s_\alpha^*(t, r_\sigma)S) \tag{131}$$

From the foregoing work, we have that if $\frac{\text{Tr}P\sigma'}{\text{Tr}\sigma'} = r_\sigma$, then $\text{Tr}\sigma'(t) = \text{Tr}\sigma'$. In particular, $\text{Tr}\sigma'(t)$ is constant provided $\text{Tr}P\sigma' = \text{Tr}P\sigma = 0$ irrespective of the value of $\text{Tr}\sigma' \neq 0$. On the other hand, if $r_{\sigma'} \neq r_\sigma$, then $\text{Tr}\sigma'(t)$ need not remain constant. The picture that emerges involves a subset of $\mathcal{B}_+(\mathcal{H})$ consisting of members having the same $r_\sigma$, allowing us to fix the operators $W_\alpha(t)$ once and for all, thereby making them independent of the preimage $\sigma$.

A notable example is the subset of perfectly asymmetric operators, which by definition satisfy $\text{Tr}P\sigma = 0$ and which we will exclusively consider for the rest of this section. If $r_\sigma = 0$, which is equivalent to $\text{Tr}P\sigma = 0$ for non-zero $\sigma$, then expressions (128) and (130) yield respectively

$$p(t) = 1 \tag{132}$$

and

$$|m(t)| \le 1 \qquad (133)$$

If we confine ourselves to a perfectly asymmetric initial operator, the conditions (120)-(123), (132) and (133) become independent of the preimage, resulting in a map (119) that is linear, completely positive and trace preserving, $\text{Tr}\sigma(t) = \text{Tr}\sigma$.

With a simpler version of Eq. (127), viz.,

$$\text{Tr}P\rho(t) = m(t), \qquad (134)$$

which is valid for a perfectly asymmetric initial density operator, we can easily track the evolution of the symmetricity. If $m(t) \to 1$ or $m(t) \to -1$, as $t \to \infty$, a transition occurs to either a state-symmetric or state-antisymmetric density operator, respectively. For concreteness, we may provide the following two examples that lead to antisymmetric states,

$$a^2(t) = \tanh^2(\kappa t) + 1 \qquad (135)$$

and

$$s^2(t) = \text{sech}^2(\kappa t), \qquad (136)$$

and symmetric states,

$$a^2(t) = \text{sech}^2(\kappa t) \qquad (137)$$

and

$$s^2(t) = \tanh^2(\kappa t) + 1. \qquad (138)$$

In these expressions, $\kappa$ is a real parameter and $a$ and $s$ are related to the coefficients in Eq. (119) according to Eqs. (124) and (125). We may check that conditions (120)-(123), (132) and (133) are satisfied. The symmetricity is given by

$$-\tanh^2(\kappa t) \sim -1, \text{ as } t \to \infty \qquad (139)$$

for the antisymmetric case,

and

$$\tanh^2(\kappa t) \sim 1, \text{ as } t \to \infty \qquad (140)$$

for the symmetric case.

Before closing this section, we discuss entropy changes associated with the foregoing time evolution. When a state vector of a system of identical particles is symmetrized or antisymmetrized, interference or exchange terms arise [1]. For example, according to the Maxwell-Boltzmann prescription, if we let $\psi(\bar{\mathbf{r}}, \underline{\mathbf{r}})$ be the position probability amplitude for a two-particle system, the probability that one of the two particles lies in $[\bar{\mathbf{r}}, \bar{\mathbf{r}} + d\bar{\mathbf{r}}]$ and the other lies in $[\underline{\mathbf{r}}, \underline{\mathbf{r}} + d\underline{\mathbf{r}}]$ is

$$N^2 \left[ |\psi(\bar{\mathbf{r}}, \underline{\mathbf{r}})|^2 + |\psi(\underline{\mathbf{r}}, \bar{\mathbf{r}})|^2 \right] d\bar{\mathbf{r}} d\underline{\mathbf{r}} \tag{141}$$

where $N$ is a normalization constant. On the other hand, for a fermionic two-particle system with antisymmetric wave function

$$N'[\psi(\bar{\mathbf{r}}, \underline{\mathbf{r}}) - \psi(\underline{\mathbf{r}}, \bar{\mathbf{r}})] \tag{142}$$

where $N'$ is another normalization constant, we find that the corresponding probability is

$$N'^2 \left[ |\psi(\bar{\mathbf{r}}, \underline{\mathbf{r}})|^2 + |\psi(\underline{\mathbf{r}}, \bar{\mathbf{r}})|^2 - \psi^*(\bar{\mathbf{r}}, \underline{\mathbf{r}})\psi(\underline{\mathbf{r}}, \bar{\mathbf{r}}) - \psi^*(\underline{\mathbf{r}}, \bar{\mathbf{r}})\psi(\bar{\mathbf{r}}, \underline{\mathbf{r}}) \right] d\bar{\mathbf{r}} d\underline{\mathbf{r}}. \tag{143}$$

Expression (143) contains interference or exchange terms that are missing in expression (141). Since loss of interference effects is often accompanied by an increase in entropy, we might anticipate that the transition from a perfectly asymmetric state to one having a symmetricity approaching $\pm 1$ would result in a decrease in collision entropy (Renyi entropy of order two), $S_R(t) = -k \ln \mathrm{Tr} \rho^2(t)$. Let us investigate three examples using the map Eq. (119) with conditions (120)-(123), (132) and (133).

If we take a density operator that is perfectly asymmetric and operator-symmetric, then it is of the form $\rho = \frac{1}{2}\rho_A + \frac{1}{2}\rho_S$ (see Eq. (107)), and if we further assume that $\mathrm{Tr}\rho_A^2 = \mathrm{Tr}\rho_S^2$, then

$$\mathrm{Tr}\rho^2(t) = \frac{1}{2}[a^4(t) + s^4(t)]\mathrm{Tr}\rho^2. \tag{144}$$

Using condition (132) and the Cauchy-Schwarz inequality, we obtain

$$\frac{\mathrm{Tr}\rho^2(t)}{\mathrm{Tr}\rho^2(0)} \leq 2 - |\mathbf{a}(t) \cdot \mathbf{s}(t)|^2 \tag{145}$$

whence

$$S_R(t) - S_R(0) \geq -k \ln[2 - |\mathbf{a}(t) \cdot \mathbf{s}(t)|^2]. \tag{146}$$

(Unlike the notation $(P \cdot P)\rho = P\rho P$ used earlier for operators, $\mathbf{a}(t) \cdot \mathbf{s}(t)$ denotes the inner product following the physics convention of conjugating in the first variable, $\mathbf{a}(t) \cdot \mathbf{s}(t) = \sum_\alpha a_\alpha^*(t) s_\alpha(t)$.) If $\mathbf{a}(t) = \mathbf{s}(t)$, then from Eqs. (132) and (144) we obtain $S_R(t) = S_R(0)$; the entropy is constant. On the other hand, if $\mathbf{a}(t)$ and $\mathbf{s}(t)$ are perpendicular, then from Eq.(146), $S_R(t) - S_R(0) \geq -k \ln 2$ and we cannot exclude, using only the last inequality, the possibility that the entropy at a given time is less than its initial value.

For the example that we mentioned above that leads to evolution to a symmetric state, viz., $s^2(t) = \tanh^2(t) + 1$ and $a^2(t) = \text{sech}^2(t)$, and assuming that the initial density operator is perfectly asymmetric, operator-symmetric and $\text{Tr}\rho_A^2 = \text{Tr}\rho_S^2$, we compute

$$S_R(t) - S_R(0) = -k \ln\left[1 + \left(\frac{e^{2\kappa t} - 1}{e^{2\kappa t} + 1}\right)^4\right]. \tag{147}$$

On inspection, we find that the entropy is a decreasing function in the time interval $[0, \infty)$. The same Eq. (147) holds if we take the example that leads to an asymmetric state, viz., $a^2(t) = \tanh^2(t) + 1$ and $s^2(t) = \text{sech}^2(t)$.

### V. Collision Between Two Identical Particles Having Negligible Spin Interactions

In this section, we use the preceding theory to examine a collision in the center of mass frame between two identical particles of arbitrary spin $s$. The colliding particles approach one another with linear momenta $p\mathbf{e}_z$ and $-p\mathbf{e}_z$, where $\mathbf{e}_z$ is the unit vector in the $z$ direction and $p \neq 0$ is the magnitude of the linear momentum of each particle. We shall assume that the particles interact via a spin-independent potential that only depends on the distance between them. In particular, we will take for the total Hamiltonian a sum of terms in which the internal (spin) and external (non-spin) variables are uncoupled:

$$H_T = H_{\text{non-spin}} + H_{\text{spin}} \tag{148}$$

This leads to a factored total evolution operator

$$U_T(t) = \exp(-iH_{\text{non-spin}}t/\hbar)\exp(-iH_{\text{spin}}t/\hbar)$$
$$\equiv U(t)U_{\text{spin}}(t) \tag{149}$$

The spin Hamiltonian $H_{\text{spin}}$ is itself assumed to contain no coupling terms with $|m_s(1); \overline{m}_s(2)\rangle$ being a stationary state thereof,

$$H_{\text{spin}}|m_s(1); \overline{m}_s(2)\rangle = E(m_s, \overline{m}_s)|m_s(1); \overline{m}_s(2)\rangle, \tag{150}$$

$$\langle m_s'(1); \overline{m}_s'(2)|m_s(1); \overline{m}_s(2)\rangle = \delta_{m_s m_s'}\delta_{\overline{m}_s \overline{m}_s'} \tag{151}$$

and

$$[P, H_{\text{non-spin}}] = [P, H_{\text{spin}}] = 0. \tag{152}$$

The index $m_s$, which ranges from $-s$ to $s$ in increments of one, labels the $2s+1$ eigenvalues $m_s\hbar$ of the spin component along the $z$ direction. By hypothesis, the spin states of the particles are conserved throughout the collision process.

Other than knowing $s$, we may not have information about the internal quantum numbers of the particles. If the initial state is a statistical mixture of equally weighted spin states, an appropriate initial density operator is

$$\frac{1}{\sqrt{2}}(1+\varepsilon P)\rho_0 \frac{1}{\sqrt{2}}(1+\varepsilon P) \tag{153}$$

where the perfectly asymmetric state $\rho_0$ is expressed as

$$\rho_0 = \frac{1}{(2s+1)^2} \sum_{m_s,\overline{m}_s=-s}^{s} |-p\mathbf{e}_z, m_s(1); p\mathbf{e}_z, \overline{m}_s(2)\rangle\langle-p\mathbf{e}_z, m_s(1); p\mathbf{e}_z, \overline{m}_s(2)| \tag{154}$$

and $\varepsilon = 1$ or $-1$ according to whether the particles are bosons or fermions. (We are taking the liberty of calling $\rho_0$ a "state" and a "density operator" here and throughout this section even though the momemtum kets $|\pm p\mathbf{e}_z\rangle$ are not normalized and therefore cannot correspond to a physical state of a particle.)

The probability density $\mathcal{P}(-p\mathbf{n}, p\mathbf{n}, t)$ that the two particles scatter elastically in the $\pm \mathbf{n}$ directions---whatever their spin states---is then given by the following expectation value

$$\mathcal{P}(-p\mathbf{n}; p\mathbf{n}, t) = 2\mathrm{Tr} U_T(t) \frac{1}{\sqrt{2}}(1+\varepsilon P_{21})\rho_0(1+\varepsilon P_{21}) \frac{1}{\sqrt{2}} U_T^\dagger(t) \mathcal{T}_s\left(O_{[-s,s]}\right) \tag{155}$$

where the operator

$$O_{[-s,s]} = \sum_{m_s,\overline{m}_s=-s}^{s} |-p\mathbf{n}, m_s(1); p\mathbf{n}, \overline{m}_s(2)\rangle\langle-p\mathbf{n}, m_s(1); p\mathbf{n}, \overline{m}_s(2)| \tag{156}$$

A straightforward calculation leads to

$$\begin{aligned}\mathcal{P}(-p\mathbf{n}, p\mathbf{n}, t) &= |F(\mathbf{n},t)|^2 + |F(-\mathbf{n},t)|^2 \\ &+ \frac{\varepsilon}{2s+1}[F^*(\mathbf{n},t)F(-\mathbf{n},t) + F(\mathbf{n},t)F^*(-\mathbf{n},t)]\end{aligned} \tag{157}$$

where

$$F(\mathbf{n},t) = \langle-p\mathbf{n}(1); p\mathbf{n}(2)|U(t,t_0)|-p\mathbf{e}_z(1); p\mathbf{e}_z(2)\rangle. \tag{158}$$

The solution (157) agrees with the standard one, which may be found in [1], p.1448 from which we have borrowed some notation.

Eq. (155) may be rewritten as

$$\mathcal{P}(-p\mathbf{n}, p\mathbf{n}, t) = 4\text{Tr}U_T(t)\frac{1}{2}(1+\varepsilon P)\mathcal{T}_S(\rho_0)(1+\varepsilon P)\frac{1}{2}U_T^\dagger(t)O_{[-s,s]} \qquad (159)$$

The quantity $\frac{1}{2}(1+\varepsilon P_{21})\mathcal{T}_S(\rho_0)(1+\varepsilon P_{21})\frac{1}{2}$ in Eq. (159) involves projection operators that can abruptly transform a statistical mixture $\rho_0$ of factored states describing distinguishable particles to an entangled, symmetrized statistical mixture describing indistinguishable particles. (For scalar particles with $s = 0$, $\rho_0$ is pure.) In line with the theory of environment-induced symmetrization developed in the previous sections, the probability $\mathcal{P}(-p\mathbf{n}; p\mathbf{n}, t)$ becomes $\mathcal{P}_\infty(-p\mathbf{n}; p\mathbf{n}, t)$ after replacing

$$\frac{1}{\sqrt{2}}(1+\varepsilon P) \cdot (1+\varepsilon P^\dagger)\frac{1}{\sqrt{2}}$$

with

$$\sum_\alpha (a_\alpha(t)A + s_\alpha(t)S) \cdot (a_\alpha^*(t)A + s_\alpha^*(t)S)$$

and $\mathcal{T}_S$ with $e^{-\frac{\tau(t)}{2}\{P,\cdot,P\}}$ where $\tau(t)$ is a monotonically increasing function of time (in the model we looked at in Section IV, $\tau = \frac{2\eta\pi t}{\hbar^3 \beta}$),

$$\mathcal{P}_\infty(-p\mathbf{n}; p\mathbf{n}, t)$$
$$= 2\text{Tr}U_T(t)\sum_\alpha (a_\alpha(t)A + s_\alpha(t)S)\, e^{-\frac{\tau(t)}{2}\{P,\cdot,P\}}(\rho_0)(a_\alpha^*(t)A \qquad (160)$$
$$+ s_\alpha^*(t)S)U_T^\dagger(t)O_{[-s,s]}.$$

We note that $e^{-\frac{\tau(t)}{2}\{P,\cdot,P\}}(\rho_0)$ belongs to the class of density operators that are perfectly asymmetric ($\text{Tr}Pe^{-\frac{\tau(t)}{2}\{P,\cdot,P\}}(\rho_0) = 0$), and hence its norm is preserved provided we take conditions (120)-(123), (132) and (133) for $a_\alpha$ and $s_\alpha$. We further note that in the model investigated in Section IV, $e^{-\frac{\tau(t)}{2}\{P,\cdot,P\}}(\rho_0)$ becomes operator-symmetric ($Pe^{-\frac{\tau(t)}{2}\{P,\cdot,P\}}(\rho_0)P \sim e^{-\frac{\tau(t)}{2}\{P,\cdot,P\}}(\rho_0)$ as $t \to \infty$), and does so exponentially fast.

A computation reveals

$$\mathcal{P}_\infty(-p\mathbf{n}, p\mathbf{n}, t) = \left[1 + \frac{1}{2}e^{-2\tau(t)}\sum_\alpha\bigl(s_\alpha(t)a_\alpha^*(t) + a_\alpha(t)s_\alpha^*(t)\bigr)\right]|F(\mathbf{n},t)|^2$$
$$+ \left[1 - \frac{1}{2}e^{-2\tau(t)}\sum_\alpha\bigl(s_\alpha(t)a_\alpha^*(t) + a_\alpha(t)s_\alpha^*(t)\bigr)\right]|F(-\mathbf{n},t)|^2$$
$$+ \frac{1}{2(2s+1)}\bigl\{s^2(t) - a^2(t) + e^{-2\tau(t)}\sum_\alpha[a_\alpha(t)s_\alpha^*(t) - s_\alpha(t)a_\alpha^*(t)]\bigr\}F(\mathbf{n},t)F^*(-\mathbf{n},t)$$

$$+ \frac{1}{2(2s+1)} \left\{ s^2(t) - a^2(t) \right.$$

$$\left. + e^{-2\tau(t)} \sum_\alpha [s_\alpha(t)a_\alpha^*(t) - a_\alpha(t)s_\alpha^*(t)] \right\} F(-\mathbf{n},t)F^*(\mathbf{n},t). \tag{161}$$

As $t \to \infty$, we recover the standard answer (157) provided $\frac{1}{2}(s^2(t) - a^2(t)) \to 1$ or $-1$ for bosons and fermions respectively and $a_\alpha(t)s_\alpha^*(t)e^{-2\tau(t)} \to 0$.

## VI. Discussion

In the previous sections, we have developed a completely positive map that can convert perfectly asymmetric states to symmetric or antisymmetric ones. If instead we allow an arbitrary but fixed symmetricity, conditions (128) and (129) should be used instead of (132) and (133). As long as we restrict our domain to a subset of $\mathcal{B}_+(\mathcal{H})$ having a constant $r_\sigma = \mathrm{Tr}P\sigma/\mathrm{Tr}\sigma$, a completely positive, operator-sum representation of the form (119) exists since the preimage $\sigma$ enters conditions (128) and (130) only through this ratio.

To gain insight into the possibility of expanding the domain to all of $\mathcal{B}_+(\mathcal{H})$, whose members include density operators having varying symmetricities, let us rewrite Eq. (119) as

$$\Lambda_t(\sigma) = \sigma(t)$$

$$= \frac{1}{4}\left[ \begin{array}{l} (\mathbf{a}(t) + \mathbf{s}(t)) \cdot (\mathbf{a}(t) + \mathbf{s}(t))\sigma + (\mathbf{s}(t) - \mathbf{a}(t)) \cdot (\mathbf{s}(t) - \mathbf{a}(t))P\sigma P \\ +(\mathbf{s}(t) + \mathbf{a}(t)) \cdot (\mathbf{s}(t) - \mathbf{a}(t))P\sigma + (\mathbf{s}(t) - \mathbf{a}(t)) \cdot (\mathbf{s}(t) + \mathbf{a}(t))\sigma P \end{array} \right] \tag{162}$$

Eq. (128) may be expressed as

$$\frac{1}{4}(\mathbf{s}(t) + \mathbf{a}(t)) \cdot (\mathbf{s}(t) + \mathbf{a}(t))$$

$$= 1 - \frac{1}{4}(\mathbf{s}(t) - \mathbf{a}(t)) \cdot (\mathbf{s}(t) - \mathbf{a}(t)) - \frac{1}{4}[(\mathbf{s}(t) - \mathbf{a}(t)) \cdot (\mathbf{s}(t) + \mathbf{a}(t)) + (\mathbf{s}(t) + \mathbf{a}(t)) \cdot (\mathbf{s}(t) - \mathbf{a}(t))]\frac{\mathrm{Tr}P\sigma}{\mathrm{Tr}\sigma} \tag{163}$$

Inserting this last equation into Eq. (162), we obtain

$$\Lambda_t(\sigma) = \sigma + \frac{1}{4}(\mathbf{s}(t) - \mathbf{a}(t)) \cdot (\mathbf{s}(t) - \mathbf{a}(t))(P\sigma P - \sigma)$$

$$+\frac{1}{4}(\mathbf{s}(t)+\mathbf{a}(t))\cdot(\mathbf{s}(t)-\mathbf{a}(t))\left(P\sigma-\sigma\frac{\text{Tr}P\sigma}{\text{Tr}\sigma}\right)+\frac{1}{4}(\mathbf{s}(t)-\mathbf{a}(t))$$
$$\cdot(\mathbf{s}(t)+\mathbf{a}(t))\left(\sigma P-\sigma\frac{\text{Tr}P\sigma}{\text{Tr}\sigma}\right) \tag{164}$$

In view of Eqs. (129) and (130), a uniform bound for $m(t)$ is given by

$$|m(t)|\leq\frac{1}{2} \tag{165}$$

Eq. (164) is manifestly trace-preserving, at the expense of having lost manifest positivity. We can also verify that any state-symmetric or antisymmetric initial density operator remains stationary when governed by Eq. (164) (see Appendix), which implies that a state that is initially symmetric maintains this characteristic, and likewise for an antisymmetric state. Thus, this map has many nice features including preservation of the Hermiticity, positivity and trace for any $\sigma \in \mathcal{B}_+(\mathcal{H})$. However, on account of the presence of $r_\sigma$, $\Lambda_t$ is non-linear, and although positive, it is not completely positive. In fact, as we show in the Appendix, it is not even two-positive. The map $\Lambda_t$ suffers the same fate as reduced dynamics of entangled total states [30], [31]. Nevertheless, it has been recognized that maps that are not completely positive have uses in quantum applications [32], [33]. Whether the foregoing complications of Eq. (164) are acceptable may depend on the physical context.

## VII. Appendix

Here we show that although $\Lambda_t$ of Eq. (164) preserves norm (manisfestly so), Hermiticity and positivity, it is not two-positive, and therefore not completely positive. (In this paper, both a positive semidefinite operator $O$ (meaning $\langle\psi|O|\psi\rangle \geq 0\ \forall\psi$) and a positive definite operator are referred to as positive operators.)

First, we have already remarked that if Hermitian operator $\sigma$ is state-symmetric or antisymmetric then $\sigma$ is operator-symmetric [12] for if $P\sigma = \pm\sigma$, then $(P\sigma P)^\dagger = (\pm\sigma P)^\dagger \Rightarrow P\sigma P = \pm P\sigma = \sigma$. Thus, in view of Eq. (164), if Hermitian $\sigma$ is either state-symmetric or state-antisymmetric, then $\sigma(t) = \sigma$. We will use this fact later.

Now, take for instance a $2 \times 2$ density matrix describing one particle; then, two such particles are described by a $4 \times 4$ density matrix. To prove that $\Lambda_t$ is not two-positive, it suffices to find four $4 \times 4$ matrices $M_{ij}$ ($i,j = 1,2$), all elements of the domain of $\Lambda_t$, such that the $8 \times 8$ matrix $\sum_{i,j=1}^{2} E_{ij} \otimes M_{ij}$ is positive, but $\sum_{i,j=1}^{2} E_{ij} \otimes \Lambda_t(M_{ij})$ is not, where the matrix basis elements $E_{ij}$ are defined to be $2 \times 2$ matrices with a one at the $i,j$ position and a zero at the other three positions. (We use the same notation $\Lambda_t$ to denote a map of matrices, such as the matrix $M_{11}$, and a map of corresponding operators, such as the operator $\hat{M}_{11}$ introduced below.)

We want a simple counterexample, so we proceed as follows. Take $\mathbf{a}(t) \cdot \mathbf{s}(t) = 0$ and choose an initial density operator $\hat{M}_{11}$, corresponding to $M_{11}$, that satisfies $P\hat{M}_{11}P = \hat{M}_{11}$ and $\mathrm{Tr} P\hat{M}_{11} = 0$. Then Eq. (164) simplifies to

$$\Lambda_t(\hat{M}_{11}) = [1 + m(t)P]\hat{M}_{11}. \tag{166}$$

Since $\hat{M}_{11}$ is a perfectly asymmetric state that is operator-symmetric, we know that it is of the form $\frac{1}{2}(\rho_S + \rho_A)$ (cf. Eq. (107)). A concrete example is:

$$\hat{M}_{11} = \tfrac{1}{2}\left[|++\rangle\langle++| + \tfrac{1}{2}(|+-\rangle - |-+\rangle)(\langle+-| - \langle-+|)\right] \tag{167}$$

which, with the association

$$|++\rangle \leftrightarrow \begin{pmatrix}1\\0\\0\\0\end{pmatrix},\ |+-\rangle \leftrightarrow \begin{pmatrix}0\\1\\0\\0\end{pmatrix},\ |-+\rangle \leftrightarrow \begin{pmatrix}0\\0\\1\\0\end{pmatrix},\ |--\rangle \leftrightarrow \begin{pmatrix}0\\0\\0\\1\end{pmatrix},$$

corresponds to the matrix

$$M_{11} = \begin{pmatrix} \frac{1}{2} & 0 & 0 & 0 \\ 0 & \frac{1}{4} & -\frac{1}{4} & 0 \\ 0 & -\frac{1}{4} & \frac{1}{4} & 0 \\ 0 & 0 & 0 & 0 \end{pmatrix}. \tag{168}$$

Letting $\delta > 0$, we fix the other three matrices,

$$M_{ij} = E_{11} \otimes \begin{pmatrix} \delta & 0 \\ 0 & 0 \end{pmatrix} \text{ if } i \times j \neq 1. \tag{169}$$

$\Lambda_t(M_{11})$ may be computed using Eq. (166). As for the other $M_{ij}$, noting that $E_{11} \otimes \begin{pmatrix} \delta & 0 \\ 0 & 0 \end{pmatrix}$ corresponds to a Hermitian operator that is state-symmetric (viz., $\delta|++\rangle\langle++|$), we can use the invariance $\sigma(t) = \sigma$ to arrive at

$$\Lambda_t(M_{ij}) = \begin{cases} \begin{pmatrix} \frac{1}{2}[m(t)+1] & 0 & 0 & 0 \\ 0 & \frac{1}{4}[1-m(t)] & \frac{1}{4}[m(t)-1] & 0 \\ 0 & \frac{1}{4}[m(t)-1] & \frac{1}{4}[1-m(t)] & 0 \\ 0 & 0 & 0 & 0 \end{pmatrix} & \text{if } i=j=1 \\ E_{11} \otimes \begin{pmatrix} \delta & 0 \\ 0 & 0 \end{pmatrix} & \text{otherwise} \end{cases} \quad (170)$$

such that $|m(t)| \leq \frac{1}{2}$ (see expression (165)).

Exploiting the sparsity of the matrix, it is not difficult to compute that the eigenvalues of $\sum_{i,j=1}^{2} E_{ij} \otimes M_{ij}$ are $\frac{1}{2}\delta \pm \frac{1}{4}\sqrt{20\delta^2 - 4\delta + 1} + \frac{1}{4}, \frac{1}{2}, 0$, which are all non-negative if $\delta < \frac{1}{2}$. On the other hand, the eigenvalues of $\sum_{i,j=1}^{2} E_{ij} \otimes \Lambda_t(M_{ij})$ include

$$\frac{1}{4}m(t) + \frac{1}{2}\delta + \frac{1}{4} - \frac{1}{4}\sqrt{m^2(t) - 4m(t)\delta + 2m(t) + 20\delta^2 - 4\delta + 1},$$

which is negative if $m(t) < 2\delta - 1$. Both of these being Hermitian matrices, we conclude that $\sum_{i,j=1}^{2} E_{ij} \otimes M_{ij}$ is positive, but $\sum_{i,j=1}^{2} E_{ij} \otimes \Lambda_t(M_{ij})$ is not when the last two inequalities hold. The map $\Lambda_t$ is not completely positive.